\documentclass[A4paper,12pt]{article}
\usepackage[T1]{fontenc}
\usepackage[english]{babel}
\usepackage[cp1250]{inputenc}
\usepackage{epsfig}
\usepackage{amsfonts}
\usepackage{fancyhdr}
\usepackage{syntonly}
\usepackage{graphicx}

\begin{document}
\selectlanguage{english}

\begin{titlepage}
\begin{center}
\vspace*{3cm}

\begin{title}
\bold {\Huge Superposition models and the multiplicity fluctuations
in heavy ion collisions
 }
\end{title}

\vspace{2cm}

\begin{author}
\Large K. FIA{\L}KOWSKI\footnote{e-mail address:
fialkowski@th.if.uj.edu.pl}, R. WIT\footnote{e-mail address:
romuald.wit@uj.edu.pl}

\end{author}

\vspace{1cm}

{\sl M. Smoluchowski Institute of Physics\\ Jagellonian University \\

30-059 Krak{\'o}w, ul.Reymonta 4, Poland}

\vspace{2cm}

\begin{abstract}
  A class of simple superposition models based on the Glauber picture of multiple collisions is
compared with the data on the centrality dependence of the multiplicity distributions in a
central rapidity bin. We show how the results depend on the specific assumptions concerning the
distributions in the number of participants and their relations to the distributions of the number
of produced hadrons in various phase space bins. None of the versions of the model describes satisfactorily
the centrality dependence of the scaled dispersion.
\end{abstract}

\end{center}

\vspace{2cm}



\end{titlepage}

\section{Introduction}
 In a recent note \cite{KFRW} we have analyzed the PHENIX data \cite{PHE} on the
centrality dependence of the multiplicity distributions in a central
pseudorapidity bin for heavy ion collisions. We have shown that a
superposition model which presents the final state as a simple
superposition of the final states from nucleon-nucleon collisions
(described by the PYTHIA 8 generator \cite{SMS}) fails to describe
the data. Even if the average number of nucleon-nucleon
collisions is treated as a free parameter to be fitted to the
average multiplicity for each centrality bin, the dispersion values
for these bins disagree with the data.
\par
We have suggested two possible reasons for this failure. First, the
multiplicity distributions for the pp collisions measured by PHENIX
are not reproduced correctly by PYTHIA with the default parameter
values; the fluctuations in data are significantly bigger than in
the model. Second, it is well known that the average multiplicities
in heavy ion collisions are not exactly proportional to the number
of "wounded" nucleons \cite{BBC} in colliding nuclei \cite{PHO}.
Thus it does not seem reasonable to assume that the final state may
be considered as a simple superposition of final states from
nucleon-nucleon collisions.
\par
One may try to remove the deficiencies listed above taking into
account two following remarks. First, it is well known that the
multiplicity distributions in any pseudorapidity bin for the $pp$
collisions may be successfully parametrized by negative binomial
distribution \cite{UA5}. Such a distribution with the parameters
fitted to the data should be thus used instead of that from PYTHIA
with the default parameter values. Second, the wounded nucleon model
is known to work much better when modified to the "wounded quark
model" \cite{BCF}.  In a simple version of this model presented
recently \cite{BB} each nucleon consists of a quark and a diquark.
The nucleon-nucleon collision may be well approximated by the
interaction of just one component from each nucleon (both
contributing similarly to the multiplicity of the final state).
However, for the nucleon interacting more than once in the heavy ion
collision, both components are likely to interact and one gets a
double contribution to the multiplicity of the final state.
\par
Another possible improvement of the model is the inclusion of
fluctuations in the number of participants for a given impact
parameter. The reliable description of such fluctuations requires
the detailed knowledge of the nuclear structure, but one may
estimate the influence of this effect by considering simple
distributions "bracketing" the physically plausible distributions
from below and above.
\par
However, such modifications are difficult to be built into the full-fledged
Monte Carlo generator (e.g., PYTHIA). Therefore we construct a class of simple generators which,
for consecutive values of the impact parameter, produce the samples of "events" consisting
only of two numbers. These are the numbers of charged hadrons in the phase space regions
corresponding to the two detectors used in the PHENIX experiment. Such "events" are then processed
exactly in the same way as the PHENIX data.
\par
In the next section we describe the construction of
such generators, introducing gradually the improvements discussed above and comparing the consecutive
versions with the data. In the last section the results are summarized
and some conclusions are drawn.

\section {Simple models and the PHENIX data}
The PHENIX data for the multiplicity distributions in a small
central pseudorapidity bin are collected for various beams, energies
and the centrality classes defined by multiplicity ranges in the
auxiliary "BBC counters" covering the pseudorapidities $\eta$ from
the range $3<|\eta|<3.9$. As before \cite{KFRW}, we concentrate here
on the AuAu data for $200$ GeV, as the dependence on the energy and
atomic number is not very difficult to reproduce.
\par
To implement the superposition idea we use the code based on the
Glauber formalism \cite{EKL} for the heavy ion collision which
calculates for each impact
parameter the number of interacting nucleons in two colliding nuclei \cite{MIS}.
This code was already used for comparison of the wounded nucleon
and the wounded quark models \cite{EV}. In the following we assume that the
average multiplicities in the BBC counters $<n_{BBC}> $ and in the
central counter $<n_c>$ are proportional to the global number of
interacting nucleons in both nuclei $N_p$:
$$<n_{BBC}> = \alpha N_p, \quad <n_c> = \beta N_p.$$

\subsection {Wounded nucleons}
Let us assume a simple geometric distribution of the impact
parameter, in which the number of events is proportional to $b$. To
relate the generator to the experimental data we need the values of
the two constants $\alpha$ and $\beta$. To reproduce the
experimental distribution of $n_{BBC}$  presented by PHENIX
\cite{PHE2} we take $\alpha =5.2$. A fit to the dependence of
$<n_c>$ on $N_p$ shown in ref. \cite{PHE} gives the value of $\beta
= 0.18$.
\par
In the PHENIX data each centrality class (defined by the limits on
$n_{BBC}$) was labeled by the value of $N_p$ corresponding to this
range of $n_{BBC}$ in the HIJING event generator. In our model this
correspondence is only slightly different; the average values of
$N_p$ in all classes (defined by the same bounds on $n_{BBC}$)
here and in the next subsections are shifted in the worst case by a few percent. For
consistency, in what follows, we use our values of $N_p$ on the $x$
axes of the plots.
\par
The simplest assumption on the multiplicity distributions in the BBC
and central counters (for fixed $b$ and $N_p$) is to describe them
by the Poisson distribution. Then one may calculate the scaled
dispersion of the distribution in $n_c$ (for the central counter)
for each centrality class as defined by a cut in  $n_{BBC}$. More
precisely, we calculate for each value of $b$ the number of the
events $N_{ev}$ to be generated (proportional to $b$) and the values
of $N_p$,  $<n_{BBC}>$  and  $<n_c>$. Then we generate $N_{ev}$
values of $n_{BBC}$ according to the Poisson distribution with
average $<n_{BBC}>$ and register the numbers $N^i_{ev}$ of the
values falling in the ranges which define consecutive centrality
classes\footnote {An independent consistency check of our procedure
is the fact that the numbers of events in all classes defined by the
limiting values of $n_{BBC}$ are almost the same (with a few percent
accuracy); remember that these values were selected by PHENIX to
divide the global sample of events into the equally populated bins.
We repeat this check in the following for all versions of the
model.}. Finally, these values of $N^i_{ev}$ are used to decide how
many values of $n_c$ should be generated according to the Poisson
distribution for each class. The final distribution of $n_c$ for
each class is the sum of the distributions generated for all values
of $b$.
\begin{figure}[h]
\centerline{\epsfig{figure=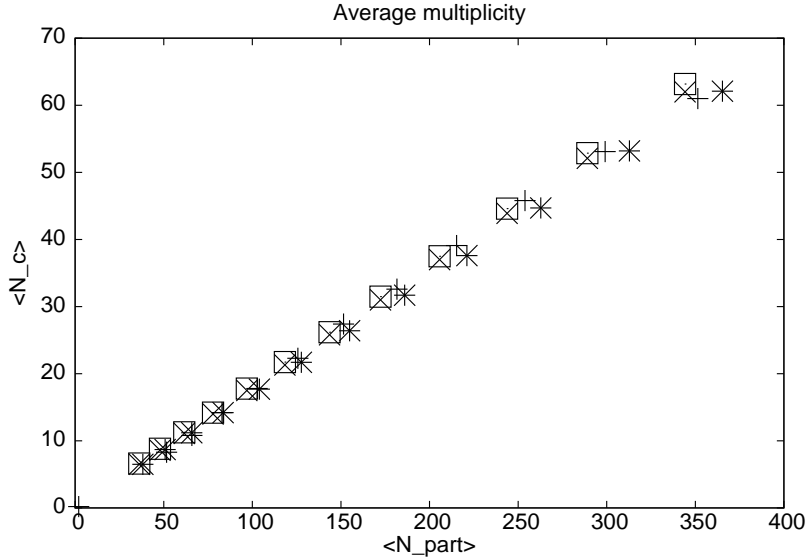,height=7.5cm}}
\caption{\footnotesize \label{Average} The average multiplicity in
the central detector for the PHENIX $pp$ and $AuAu$ data ($+$ marks)
and the superposition model ($x$ marks). The results of other
versions of the model, shown as squares and stars, will be commented
upon later}
\end{figure}
\par
This procedure by definition reproduces correctly the experimental
dependence of the value of $<n_c>$ ~on the corresponding average
value of $N_p$ defined for the consecutive centrality classes. This
is illustrated in Fig.1, where for transparency the big experimental
error bars are omitted. The real test of the model is thus the
centrality dependence of the scaled variance
$$\omega = (<n_c^2>-<n_c>^2)/<n_c>.$$
The results are shown and compared with the PHENIX data in Fig.2
together with the results from two other versions of the model (to
be discussed later). Here and in the following the errors are
comparable with the size of the data points.

\begin{figure}[h]
\centerline{ \epsfig{figure=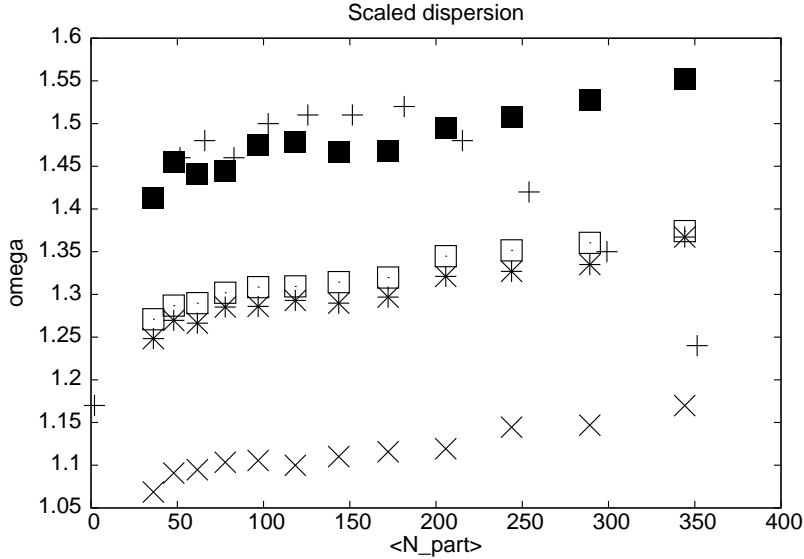,height=7.5cm}}
\caption{\footnotesize \label{Omega} The scaled dispersion for the
PHENIX $pp$ and $AuAu$ data ($+$ marks) and the superposition model
with Poisson distributions ($x$ marks) and NBD distributions with
three choices of the $k$ parameters (stars, open and full squares)}
\end{figure}

We see that the fluctuations are strongly underestimated and
increase monotonically with $N_p$ in disagreement with the data.
In fact, the values are quite similar (although slightly lower) to
the results obtained by superimposing the $pp$ events generated from
PYTHIA \cite{KFRW}. This is so because the Poisson distribution is only
slightly narrower than the distribution predicted by PYTHIA for the
$pp$ collisions.
\par
In the next step we replace the Poisson distribution by the
negative binomial distribution (NBD):
$$P_{NBD}^{<n>,k}(n)=\frac{\Gamma(n+k)}{\Gamma(k)\Gamma(n+1)}\Big(\frac{<n>}{<n>+k}\Big)^n\Big(\frac{k}{<n>+k}\Big)^k$$
with parameters fitted to the PHENIX $pp$ data. More precisely, if we assume that
a single participant nucleon yields (for some part of the phase space) a charged hadron
distribution described by the NBD with the parameters $<n>$ and $k$, we expect for
the $pp$ collisions the NBD with parameters $2<n>$, $2k$, and for a heavy ion
collision with $N_p$ participants the NBD with parameters $N_p<n>$ and $N_pk$. The rest of the
procedure remains unchanged.
\par
There is an uncertainty connected with this prescription for the NBD
parameters. The PHENIX $pp$ data for the central bin are collected
on condition that there is at least one charged hadron in each of
the two BBC counters. Thus we cannot assume that the exact values of
the measured parameters of this distribution should be used to
predict the distribution for given value of $N_p$, where obviously
not all the participants must produce the hadrons falling into these
counters. Therefore in the following we compare the results for
different values of the parameters for the "elementary" NBD in $n_c$
(and $n_{BBC}$).
\par
We have performed the calculations with the average
multiplicities of the $n_c$ and $n_{BBC}$ distributions calculated
for each $N_p$ as before and with three different choices of the $k$ parameter. In
the first calculation, for all values of $N_p$ we have taken approximately the
same ratio of $k_c$ to $<n_c>$ as in the PHENIX $pp$ data (about
$6$) and assumed that for the distribution of $n_{BBC}$ this ratio
is the same. In the second one, we assumed much broader distribution of
$n_{BBC}$, with $k/<n> \sim 1$. Finally, we have chosen the distribution of $n_c$
with the ratio of $k_c$ to $<n_c>$ smaller than in the PHENIX $pp$ data by a factor
of $0.5$. The last two choices may be regarded as maximizing the fluctuations due to the spread of multiplicities
which define the centrality classes. In fact, the third choice gives such a broad distribution of $n_c$
for a given $N_p$ that it may be regarded as breaking the basic assumption of our class of models:
building the heavy ion collisions from the elementary pp collisions. Indeed, $k_c/<n_c> \simeq 3$
corresponds to the value of $\omega -1$ twice as big as in the PHENIX $pp$ data.
\par
The first two choices give very similar results, significantly
higher than the Poisson distributions, but the fluctuations in heavy
ion collisions are still underestimated (see Fig.2). The scaled
dispersion increases monotonically with the number of participants,
contrary to the data. For the third choice we match the data for a
low number of participants, but not for the central events, where
the model results keep increasing while the data fall down.

\subsection {Wounded quarks}
In this subsection a more refined model based on the ''wounded quark'' idea
\cite{BCF}, \cite{BB} is developed. We shall modify the code \cite{MIS} which was used to calculate
for each value of the impact parameter $b$ the global number
of participants  in two colliding nuclei
\begin{eqnarray}
\label{more2} N_p(\vec b) = \int d^2s T_A(\vec s) \Big\{ 1 -  \Big[1
- \frac{\sigma_{NN}T_B(\vec s - \vec b)}{B}\Big]^B\Big\} + \int d^2s
T_B(\vec s) \Big\{1 - \Big[ 1- \frac{\sigma_{NN}T_A(\vec s - \vec
b)}{A}\Big]^A\Big\}.\nonumber
\end{eqnarray}
Similar arguments allow to calculate the global number of nucleons which interacted exactly once
\begin{eqnarray}
\label{oneparticipant}
N_p^1(\vec b) = \int d^2s T_A(\vec
s) \sigma_{NN}T_B(\vec s - \vec b) \Big[ 1-
\frac{\sigma_{NN}T_B(\vec s - \vec b)}{B}\Big]^{B-1}  + \nonumber\\
+\int d^2s T_B(\vec s) \sigma_{NN}T_A(\vec s - \vec b) \Big[ 1-
\frac{\sigma_{NN}T_A(\vec s - \vec b)}{A}\Big]^{A-1}\nonumber
\end{eqnarray}
and the number of those, which interacted at least twice
\begin{eqnarray}
\label{more2} N_p^2(\vec b) = \int d^2s T_A(\vec s) \Big\{ 1 -
\Big[1 - \frac{\sigma_{NN}T_B(\vec s - \vec b)}{B}\Big]^B -
\sigma_{NN}T_B(\vec s - \vec b)\Big[1 -\frac{\sigma_{NN}T_B(\vec s -
\vec b)}{B}\Big]^{B-1}\Big\}+\nonumber \\
+ \int d^2s T_B(\vec s) \Big\{1 - \Big[ 1- \frac{\sigma_{NN}T_A(\vec
s - \vec b)}{A}\Big]^A  -\sigma_{NN}T_A(\vec s - \vec b)\Big[1 -
\frac{\sigma_{NN}T_A(\vec s - \vec b)}{A}\Big]^{A-1} \Big\}\nonumber
\end{eqnarray}
with the obvious condition $N_p=N_p^1+N_p^2.$ Assuming that the
multiple interaction results in doubling the average multiplicity
and the value of the $k$ parameter one gets for each value of $b$ the multiplicity distribution given by
$$P(n)=\gamma P_{NBD}^{N_p^1<n_0>,N_p^1k}(n)+(1-\gamma)P_{NBD}^{2N_p^2<n_0>,2N_p^2k}(n),$$
where $\gamma=N_p^1/(N_p^1+N_p^2)$. Note that the average
multiplicity in this distribution is
$$<n>=[\gamma N_p^1+2(1-\gamma )N_p^2]<n_0>.$$
To reproduce the dependence of $<n_{BBC}>$ and $<n_c>$ on $<N_p>$ we
needed the values of $<n_0>$ of $2.7$ and $0.095$, correspondingly.
The ratio $k/<n_0>$ was assumed about $6$, as before. Now we
generate the distributions in the BBC and central counters according
to this distribution and proceed as before. The same procedure is
also performed for the Poisson distribution. The dependence of the
average multiplicity on $N_p$, shown in Fig.1 as squares, is exactly
the same as in the previous version of the model (shown as the
$x$-signs). The results for the scaled dispersion are shown in
Fig.3. The values are slightly lower, but the main feature of Fig.2
remains unchanged: the dispersion increases monotonically with the
number of participants, in disagreement with the data.

\begin{figure}[h]
\centerline{ \epsfig{figure=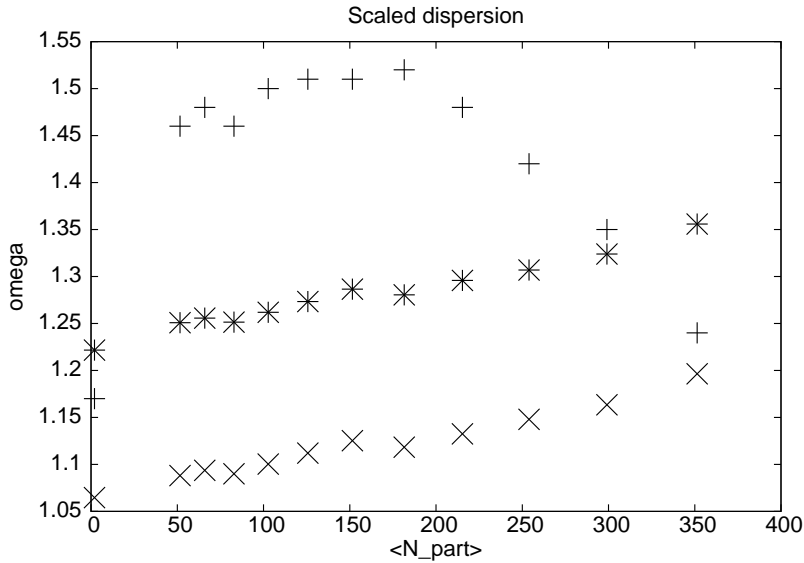,height=7.5cm}}
\caption{\footnotesize \label{Omega1} The scaled dispersion for the
PHENIX $pp$ and $AuAu$ data ($+$ marks) and the ''wounded quark''
model with the Poisson distributions ($x$ marks) and NBD
distributions (stars)}
\end{figure}

\par
Thus, contrary to our speculations formulated in \cite{KFRW}, the replacement of the "wounded nucleon" by the "wounded quark" model
did not improve the situation. The reason is that the fluctuations of $n_c$ for a given centrality class
result mainly from the fluctuations in the number of sources contributing to the values of $n_{BBC}$ in the
range defining this class. It does not really matter if these sources are nucleons or quarks (diquarks); for the
linear relation between their number and average value of $n_{BBC}$ the results are quite similar.
The slight reduction of the dispersion values is probably a consequence of the fact that the number of sources is now bigger.
\par
The values of the scaled dispersion may be increased by decreasing
the $k_c/<n_c>$ value, as in the previous subsection. However, the
centrality dependence remains monotonic in disagreement with the
data.

\subsection {Fluctuating number of participants}

There is still one assumption in the simple model we use which may be modified without abandoning the basic
superposition idea. Using the code \cite{MIS} one can calculate a well defined
value of $N_p$ (or $N_p^1$ and $N_p^2$) for each value of $b$. Obviously, in the real collision of nuclei this number may fluctuate and
the calculated value should be treated rather as an average. Moreover, there is a simple reason why such fluctuations
should not increase monotonically with centrality: for most central collisions almost all nucleons are ''wounded''
and the average is close to the maximal possible value. This may damp the fluctuations and eventually may lead to non-monotonic
dependence of the scaled dispersion on the number of participants.
\par
 In what follows we consider different distributions which may be assumed for $N_p$ around this average. The reliable choice
 would require a thorough knowledge of the nuclear structure, and in particular of the two nucleon correlations in the
 position space. However, one may use simple lower and upper limits to the fluctuations. As the first one, we use a
 simple binomial distribution of the number of participants with the average $<N_p>$ calculated from the code \cite{MIS}; its
 maximal value is obviously given by $2A$. Let us remind here that the dispersion squared of this distribution is given by
 $$D^2=<N_p>(1-<N_p>/2A).$$

\begin{figure}[h]
\centerline{ \epsfig{figure=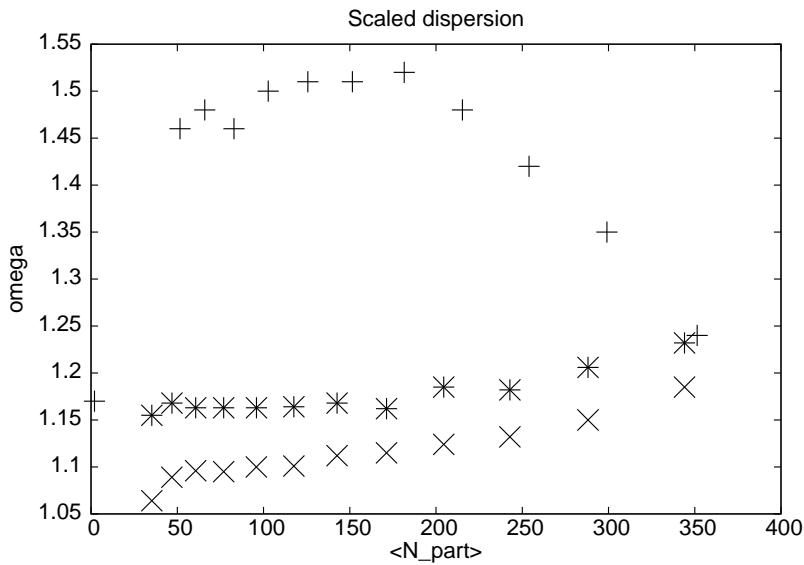,height=7.5cm}}
\caption{\footnotesize \label{Omega} The scaled dispersion for the
PHENIX $pp$ and $AuAu$ data ($+$ marks) and the superposition model
with the binomial smearing in the number of participants and the
Poisson distributions ($x$ marks) and the NBD distributions (stars)}
\end{figure}

\par
 Now the procedure is more involved. For each value of the impact parameter $b$ we calculate $<N_p>$ as before, then generate randomly the
 values of $N_p$ from the binomial distribution with this average, calculate the average values of $n_{BBC}$ and $n_c$
 (proportional to $N_p$) and generate the values of these variables according to the Poisson or NBD distributions. The results are shown in Fig.4.
\par
Contrary to the na\"{i}ve expectations, the fluctuations in the number of participants did not increase the values of
the scaled dispersion. In fact, the values for the NBD distributions are significantly lower than in Fig.2 for the same values of the $k$ parameter.
This may be interpreted as the result of the weaker correlation between the range of $n_{BBC}$ and of $n_c$, since these two values
are generated independently and the range of $N_p$ is now broader for each ''class of centrality''.
\par
The second choice, maximizing the spread of $N_p$ for a given $b$, is just a flat distribution. More precisely, we assume that for the $<N_p>$ value
below $A/4$ the probability of any value of $N_p$ between $0$ and $2<N_p>$ is the same, for the $<N_p>$ value between $A/4$ and $3A/4$ the allowed
range is between $<N_p>-A/4$ and $<N_p>+A/4$ and for $<N_p>$ above $3A/4$ the allowed range of $N_p$ is between $2A-2<N_p>$ and $2A$. Now the
dispersion squared is given in these three ranges of $<N_p>$ by
$$2<N_p>(2<N_p>+1)/6,~~~A(A/2+1)/24,~~~(2A-<N_p>)(4A-2<N_p>+1)/6.$$

\begin{figure}[h]
\centerline{ \epsfig{figure=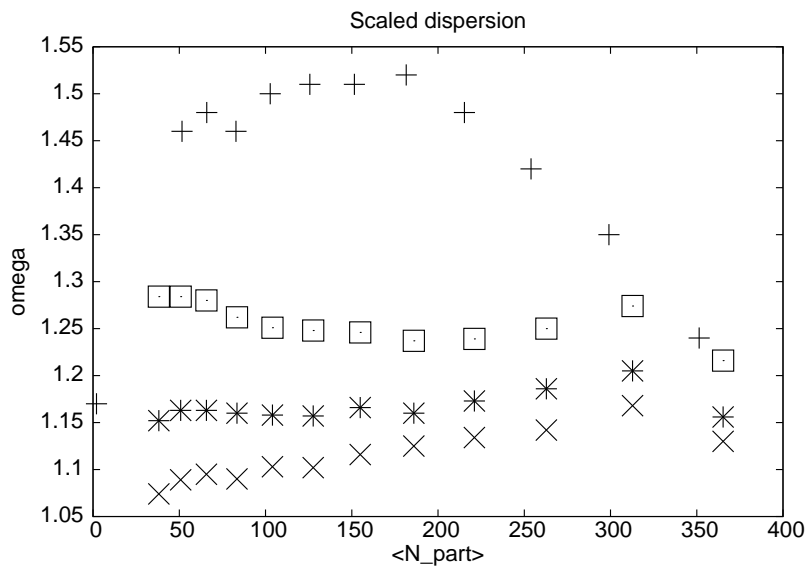,height=7.5cm}}
\caption{\footnotesize \label{Omega} The scaled dispersion for the
PHENIX $pp$ and $AuAu$ data ($+$ marks) and the superposition model
with the flat smearing in the number of participants and the Poisson
distributions ($x$ marks) and the NBD distributions (stars and
squares)}
\end{figure}

\par
These values are obviously much higher than in the case of the
binomial distribution considered before. Repeating the procedure
described above we get the scaled dispersion as shown in Fig.5. The
results for the average multiplicity are shown as stars in Fig.1.
\par
For the Poissonian distribution and for the NBD with realistic values of $k/<n>$ the results are almost the same as in Fig.4. If we choose
the value of $k/<n>$ much lower than in the $pp$ data, as already used in Fig.2, we get non-monotonical dependence of $\omega$ on $<N_p>$.
However, the shape disagrees with the data and the values are still much too low. Thus we conclude that by increasing the fluctuations in the
number of participants we are not able to reproduce the data.

\section{Conclusions and outlook}
We considered the multiplicity distributions in the central rapidity bin in the heavy ion collisions for various "centrality classes"
defined by the multiplicity in another rapidity bin. A class of simple generators is constructed basing on the assumption that the final state
is a superposition of states obtained separately from each participant nucleon. We considered the modifications resulting from
counting the "wounded quarks" instead of the "wounded nucleons" and from the fluctuations in the number of participant nucleons for the given
value of the impact parameter. None of the versions of the superposition model considered is compatible with the data. In particular,
the observed slow increase of the scaled dispersion at moderate centralities and the decrease for the most central events is not reproduced.
\par
The reason why the modifications introduced did not yield the expected improvement seems to be the particular definition of "centrality classes" used in
the PHENIX experiment. Both the "wounded quark" idea and the fluctuations in the number of participants may give saturation (or even decrease) of the
fluctuations for highest centrality assuming that the centrality is defined by a fixed range of the number of participants. However, when it is defined
by the multiplicity of hadrons in some detector (as in the PHENIX data) these expectations are not justified, as we have seen in the model calculations.
\par
Our conclusions are that it seems to be difficult (if not impossible) to describe the dependence of the scaled dispersion on centrality measured by PHENIX
in the framework of superposition models. This suggests the presence of some collective effects in these data.

\end{document}